\documentclass[letterpaper]{article} 
\usepackage[]{aaai24}  
\usepackage{times}  
\usepackage{helvet}  
\usepackage{courier}  
\usepackage[hyphens]{url}  
\usepackage{graphicx} 
\usepackage{natbib}  
\usepackage{caption} 
\usepackage{caption}
\usepackage{subcaption}
\usepackage{tikz}
\usetikzlibrary{calc}
\usepackage{algorithm}
\usepackage{algorithmic}

\usepackage{newfloat}
\usepackage{listings}
\DeclareCaptionStyle{ruled}{labelfont=normalfont,labelsep=colon,strut=off} 
\floatstyle{ruled}
\newfloat{listing}{tb}{lst}{}
\floatname{listing}{Listing}
\title{15 Years of Algorithmic Fairness\\Scoping Review of Interdisciplinary Developments in the Field}
\author{
    Daphne Lenders \& Anne Oloo
}
\affiliations{
    University of Antwerp, Antwerp, Belgium\\

}

\usepackage{bibentry}

\begin{document}

\maketitle

\begin{abstract}
This paper presents a scoping review of algorithmic fairness research over the past fifteen years, utilising a dataset sourced from Web of Science, HEIN Online, FAccT and AIES proceedings. All articles come from the computer science and legal field and focus on AI algorithms with potential discriminatory effects on population groups. Each article is annotated based on their discussed technology, demographic focus, application domain and geographical context\footnote{The data is available at \url{https://github.com/calathea21/algorithmic_fairness_scoping_review}}. Our analysis reveals a growing trend towards specificity in addressed domains, approaches, and demographics, though a substantial portion of contributions remains generic. Specialised discussions often concentrate on gender- or race-based discrimination in classification tasks.
Regarding the geographical context of research, the focus is overwhelming on North America and Europe (Global North Countries), with limited representation from other regions. 
This raises concerns about overlooking other types of AI applications, their adverse effects on different types of population groups, and the cultural considerations necessary for addressing these problems. With the help of some highlighted works, we advocate why a wider range of topics must be discussed and why domain-, technological, diverse geographical and demographic-specific approaches are needed. This paper also explores the interdisciplinary nature of algorithmic fairness research in law and computer science to gain insight into how researchers from these fields approach the topic independently or in collaboration. By examining this, we can better understand the unique contributions that both disciplines can bring.\looseness=-1
\end{abstract}

\section{Introduction}
Research on algorithmic fairness has been present for about 15 years. What initially started as a slow movement has become a popular and prominent research field, with dedicated conferences about the topic, like ACM FAccT and AAAI AIES. 
Throughout these years, the field has kept evolving, fueled by public discourses about unfair algorithms, new legislations around AI and ever-emerging technologies. While it is generally well known that the field develops rapidly, less is understood about how it has developed, what the most prominent research areas are and where the research efforts come from. Yet, only when zooming out and having a better view of the large body of literature that already exists, we get an idea of whether the research has kept up with the pace in 
of technology and where the biggest research research gaps and opportunities lay.
For this purpose, we have conducted a scoping review on the field of algorithmic fairness. Using four scientific databases, namely, Web of Science, Hein Online, ACM FAccT and AAAI AIES proceedings, we have sampled a total of 1570 papers dealing with this topic and have annotated them in terms of the domain they consider, the demographic groups they focus on and technology they discuss. 
By providing aggregated results over these three metrics, we sketch an overview of the most prominent research areas within the field, and how these have developed over the years. In doing so, we also differentiate between the research efforts coming from primarily Computer Science and Law based perspectives. We highlight how authors with different expertise approach research areas differently, and which areas remain under-addressed by either or both communities. By then highlighting some research studies in less popular areas of the field, we emphasize which areas need to be addressed to tackle algorithmic discrimination in all of its forms, rather than limited to a narrow set of technologies and domains. To summarize, the first part of our work addresses the following research questions:

\textbf{\textit{RQ1}}: How has algorithmic fairness literature developed in terms of the domains they address and what are the opportunities/gaps in adopting domain-specific approaches from a technological and legal perspective?

\textbf{\textit{RQ2}}: How has algorithmic fairness literature developed in terms of the demographic groups they focus on? How does this differ between researchers with technological and legal expertise?

\textbf{\textit{RQ3}}: How has algorithmic fairness literature developed in terms of the technologies they address? How does this differ between researchers with technological and legal expertise?

Our last research concerns the geographical context of the research on algorithmic fairness, both in terms of the authors' affiliations and the geographical areas they address. We showcase how much of the current literature is primarily centred around Global Northern countries and highlight how more recent contributions, focussing on other geographical areas, bring to light important considerations around algorithmic fairness that should not be overlooked. Hence the last research question of this study is:

\textbf{\textit{RQ4}}: What is the geographical scope of algorithmic fairness literature, both in terms of researchers' geographical affiliation and the content of their papers?

\section{Related Literature \& Motivation}
There are many literature reviews available related to algorithmic fairness. Different from scoping reviews, these works dive into specific aspects of the topic, like bias mitigation methods for classification algorithms \cite{hort2023bias}, datasets commonly used in experiments \cite{fabris2022algorithmic}, or fairness concerns related to specific technologies like computer vision system. \cite{malik2019deep}. Their goal is to summarize the most important contributions and insights surrounding these topics and identify concrete research gaps related to them.
In comparison, scoping reviews on algorithmic fairness are much more sparse. Rather than summarizing the literature on one concrete topic, scoping reviews aim to give a high-level overview of broad and general research areas that encompass many different technologies, domains and disciplines. Scoping revies aim to sketch the breadth of these areas and identify the most popular research directions. In doing so, they also highlight which areas are currently underexplored and need more attention from the research community.

Vilaza et al. (\citeyear{nunes2022scoping}) report a scoping review on ethics in technology and inspect 129 papers coming from the SIGCHI conferences. In particular, they assess the themes of the ethical considerations in each paper (e.g. privacy, discrimination, mental well-being etc.), the population groups that are discussed, and the type of technologies inspected (e.g. web applications, social media, etc.).
Similarly, a study by Birhane et al. (\citeyear{birhane2022forgotten}) dives into the topic of AI ethics across FAccT and AIES papers. They aggregate results of 535 papers, focusing on how concrete or abstract each work of literature is regarding the ethical aspects they address. In particular, they inspect whether papers discuss case studies of algorithmic systems already used by industry, and how much effort the works put into understanding how real stakeholders are affected by these systems. 
A study that emphasizes geographical regions/contexts in which AI ethics are addressed is conducted by Urman et al (\citeyear{urman2024mapping}). Specifically, they inspect 200 papers describing AI auditing studies, not just identifying which ethical aspects the AI systems are audited for, but also highlighting the countries on which the audits were focused, and the geographical affiliation of the authors contributing to these studies. 
Our contribution sets itself apart from these already existing scoping reviews in various ways:
\begin{enumerate}
    \item Different from other studies, we focus on algorithmic fairness as one sub-area of AI ethics, rather than AI ethics in general. This allows us to identify the research landscape and gaps more specific to this area, addressing the research focus in terms of addressed domains, demographic groups, technologies and geographical context
    \item We are the first scoping review, to inspect the development of the research area from an interdisciplinary perspective, focusing on how authors with Computer Science and Law expertise address this topic differently, and where the research gaps in either or both of the fields lay
    \item To the best of our knowledge our study is the largest scoping review on AI ethics, aggregating the results of a total of 1570 papers. By not merely focussing on contributions coming from FAccT and AIES, we get a better overview of the current literature.
\end{enumerate}

\section{Methodology}
To conduct our scoping review we adopt the PRISMA (short for: ``Preferred Reporting Items for Systematic Reviews and Meta-Analyses") guidelines \cite{liberati2009prisma}. This means that our methodology consisted of three key steps: the first, was devising a search strategy, by selecting the scientific databases for locating relevant papers and designing queries to search these databases. The second step was going through the found papers and deciding which ones to include in this review. The third and last step was annotating the selected papers for relevant information and analysing the results. We are going to describe each step in more detail in the following sections.

\subsection{Databases \& Search Query}
\begin{figure*}[t]
    \centering
    \includegraphics[width=0.7\textwidth]{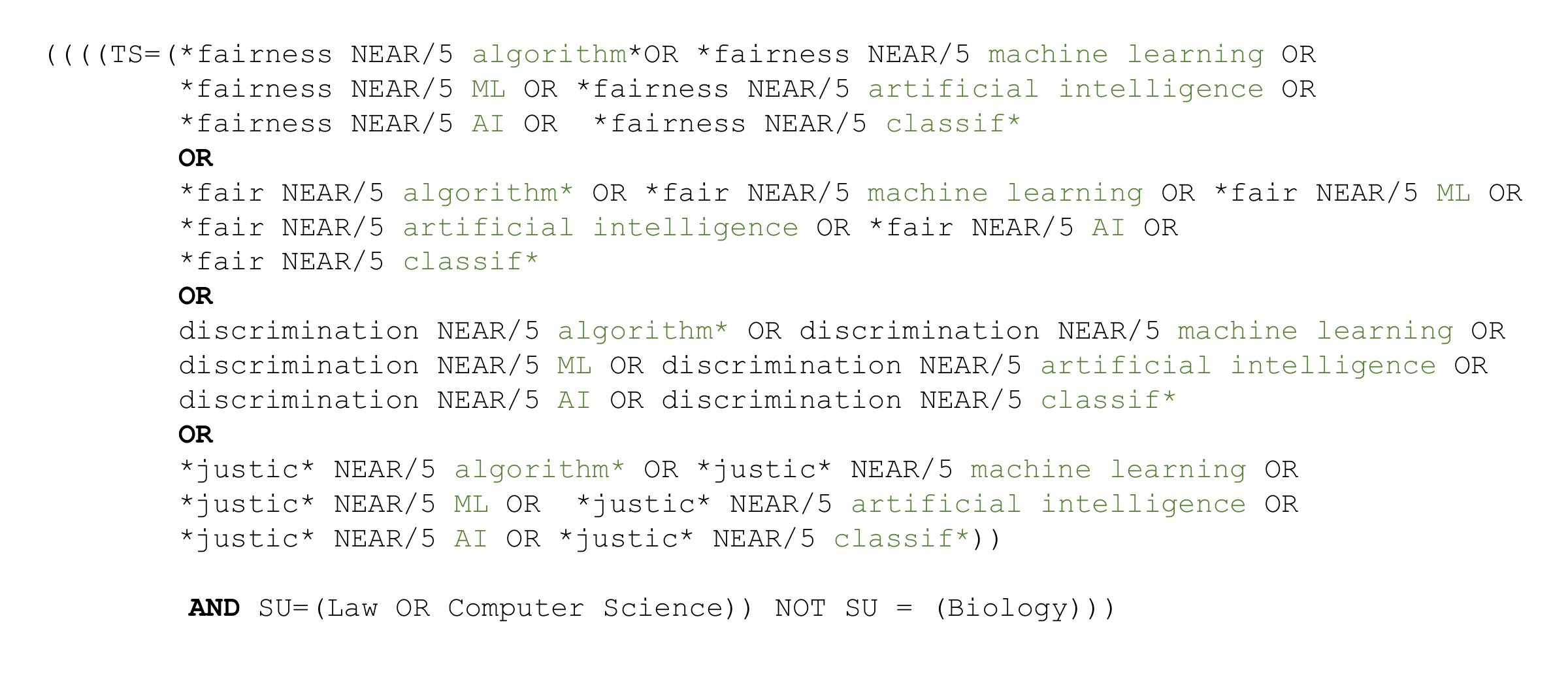}
    \caption{The Web of Science search query to capture relevant literature, based on key phrases in papers' title and abstract}
    \label{fig:wos_search_query}
\end{figure*}

We used Web of Science as our main database for scientific articles. Using their advanced search function, we set up the search query as seen in Figure \ref{fig:wos_search_query} to find papers related to algorithmic fairness, with a focus on Computer Science or Law. 
The search query uses filters to scan through papers based on their title and abstracts. It looks for specific keyword combinations in either of them. The keyword combinations are all variations of terms like “algorithmic fairness”, “fair Machine Learning”, or “discrimination in AI”. By including the wildcard operator (*), we ensured that variations of words are captured that come from the same root (e.g., including the wild card operator before and after ‘fair’ we automatically include terms like ``unfair" and ``fairly"). Further, we use the (NEAR\textbackslash5) operator to specify that two words should be placed within a distance of 5 words in the text. The search query was based on an iterative process, adding or removing terms depending on how many search results we obtained. For instance, initially, the query accounted for terms like “bias in Machine Learning”. However, as “bias” is also a purely mathematical (and not ethical) related concept, this yielded too many results, and we excluded this term. After finalising our search query, we conducted a sanity check to ensure that it captured highly cited and well-known papers.
We used variations of the same query for the database of papers from ACM FAccT and AAAI AIES proceedings, as well as Hein Online. We chose the first two, as they are the the most prominent conferences on ethics in socio-technical systems. We chose the latter because it is a database containing mostly legal sources, underrepresented in the results of Web of Science.

\subsection{Selection of Papers}
Once we executed our initial search query, we received a total of 6027 papers that required screening for their relevance to the topic of algorithmic fairness. To perform the screening, we utilised Rayyan.ai and established various inclusion and exclusion criteria.
To be included in the review, sources were required to have an abstract to ensure that each source under consideration had a minimum level of information available. 
Moreover, several categories of sources were excluded from the outset. These included introductory notes, book reviews and tutorials, as they were not expected to provide in-depth research content and were not aligned with the intended study scope. Additionally, abstracts of workshops and tutorials for which the full article or chapter could not be accessed were excluded. Lastly, language was an exclusion criterion, with sources not in English being excluded.
We then used the articles' titles as primary indicators of their relevance to the field of algorithmic fairness. In case of ambiguity, we also used the papers' abstracts to decide on their relevance. Through this selection process we ended up with a total of 1570 sources to be included in this scoping review.

\subsection{Data Extraction}\label{sec_annotation}
For each of the papers we included for this analysis several features were available, namely their title, abstract and year of publication. Many papers also had a DOI available, which we used to automatically extract additional information from them using \texttt{pybliometrics} \cite{ROSE2019100263}. This python library utilizes an API to extract information from the Scopus database. In our case, we extracted the names of the papers' authors, and for each author their affiliation at the time of writing the paper (consisting of the name of their institution as well as the corresponding country). This information would be used for answering \textit{RQ4}. To answer parts of research questions 1-3, we also extracted the main expertise areas of each author, as they had self-reported in Scopus.  

To extract information on authors' affiliation and expertise on papers without a DOI, we carried out a manual labelling process. We manually checked the papers to extract the authors' names and their affiliations at the time of writing. To determine their area of expertise, we used platforms such as Google Scholar, LinkedIn, and Research Gate. It is important to note that the different labelling processes of authors' expertise may have introduced some errors or biases in our final dataset of papers. This is because the authors' self-reported areas of expertise may differ from the ones we could establish ourselves through a basic web search. Therefore, any results that pertain to this aspect should be regarded as a proxy. Further, many papers had authors coming from mixed backgrounds, with at least one author listing both ``Computer Science" and ``Law" as their main expertise. Though generally, it could be interesting to inspect contributions from authors with such mixed backgrounds, our result analysis focuses on the work coming only from Law or only Computer Science expertise. This choice was made, because we found many of the ``mixed" expertise labels to not be completely reliable, i.e. we found that a lot of Computer Scientists listed ``Law" as one of their backgrounds, mostly because ``algorithmic fairness" is a topic with some legal implications, not because their work specifically deals with any specific legislation or other legal considerations. 

To analyse papers' domain-, demographic and technological focus, as we address in RQ1-RQ3, we manually annotated papers according to these characteristics, using their titles and abstract. We acknowledge that reading full papers would yield more precise results, but since our database consisted of more than 1500 papers, time constraints did not allow this.
Through an iterative process, we identified recurring themes regarding the three dimensions and merged similar categories into broader ones, where needed. For example, to describe the papers' technological focus we first had a separate category for ``Face Recognition", but because not many papers focussed on this topic, we decided to include them in the broader category "Computer Vision``. 

Below we list the annotation labels we ended up using for each of the three papers' features:

\begin{itemize}
    \item \textbf{Domain} - Criminal Justice, Education, Employment, Finance, Health, Judicature, Public Sector, Other, None
    \item \textbf{Demographic Groups as Based on} - Age, Disability, Gender, Intersectional, Race, Other, None
    \item \textbf{Technological Approach} - Computer Vision, Data Collection, Hybrid Human-AI, NLP, Resource Allocation, Social Networks, Unsupervised Learning, Ranking, Recommendation, Classification, Other, General
\end{itemize}

In the result analysis it will become clear that a lot of our found papers do not focus on a specific domain or demographic group (as denoted by the ``None" label for either of both features). It is important to note, that both features were only assigned a ``non-None" label if papers made some demographic group or some domain the specific focus of their research. To exemplify, many papers introduce novel bias mitigation methods for classification tasks and test their method among others on the COMPAS dataset. Even though this dataset falls under the criminal justice domain, these papers were not tagged as such, unless they specified in their abstract that they went beyond the general benchmark evaluation on this dataset, e.g. consulting domain experts' opinions on the matter or considering domain-specific legislation.
Similarly, many papers consider ``sex" or ``race" as sensitive attributes in their experimental settings. Again, their demographic focus was not tagged as such, unless they dived into specific, historically- or culturally grounded discrimination of those groups. 
Regarding the technological approach of papers, the ``General" label was used if a paper provided a literature review on algorithmic fairness or discussed this as a broad phenomenon, considering many different algorithmic approaches. Also, if a paper's technological approach fell into multiple categories, we chose the more specific one as the primary focus. For instance, a paper on hate speech classification was labelled as "Natural Language Processing" instead of "Classification".

Lastly, to provide labels for the geographical content of papers to answer RQ4, we checked if they mentioned any specific region (``Europe") or country (e.g. ``United States") in their title or abstracts and annotated them accordingly.

\section{Results}
After selecting and annotating our papers we conducted the analyses to answer our research questions as outlined in the Introduction of this paper. 
In the following sections, we will describe the results of these analyses. For each research question, we will first provide a high-level overview of the results, highlighting the research trends related to the specific areas. After, we highlight some specific case studies belonging to less popular research areas, emphasizing the need to dedicate more research to them.

\subsection{\textbf{RQ1}: The Need for Domain-Specific Approaches} \label{sec_results_domain}

\begin{figure*}[t]
    \centering
    \begin{minipage}[b]{0.52\textwidth}
        \centering
        \includegraphics[width=\textwidth]{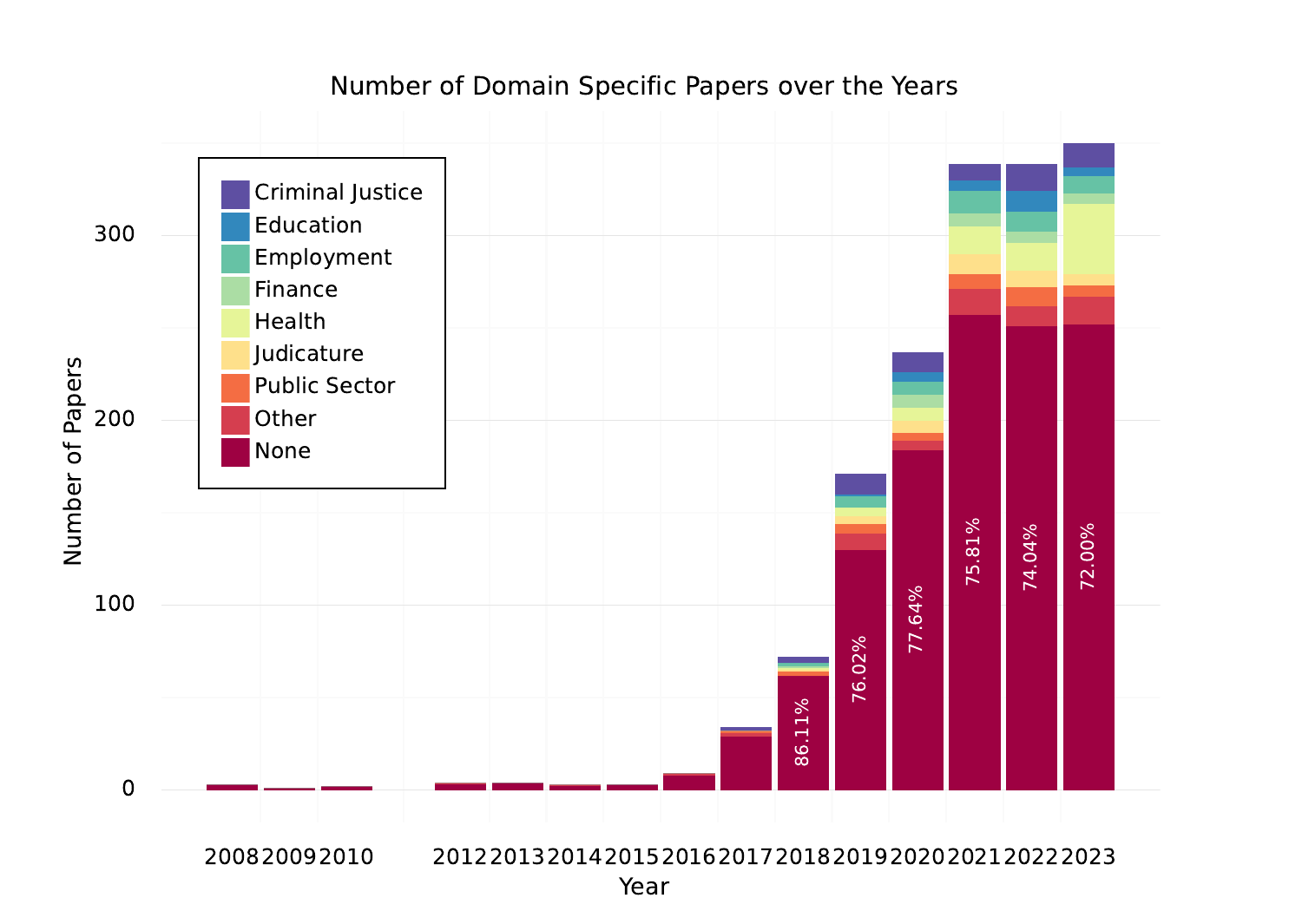}
        \caption{The domain focus of papers over the years}
        \label{fig:domain_specific_over_years}
    \end{minipage}
    \hfill
    \begin{minipage}[b]{0.45\textwidth}
        \centering
        \includegraphics[width=\textwidth]{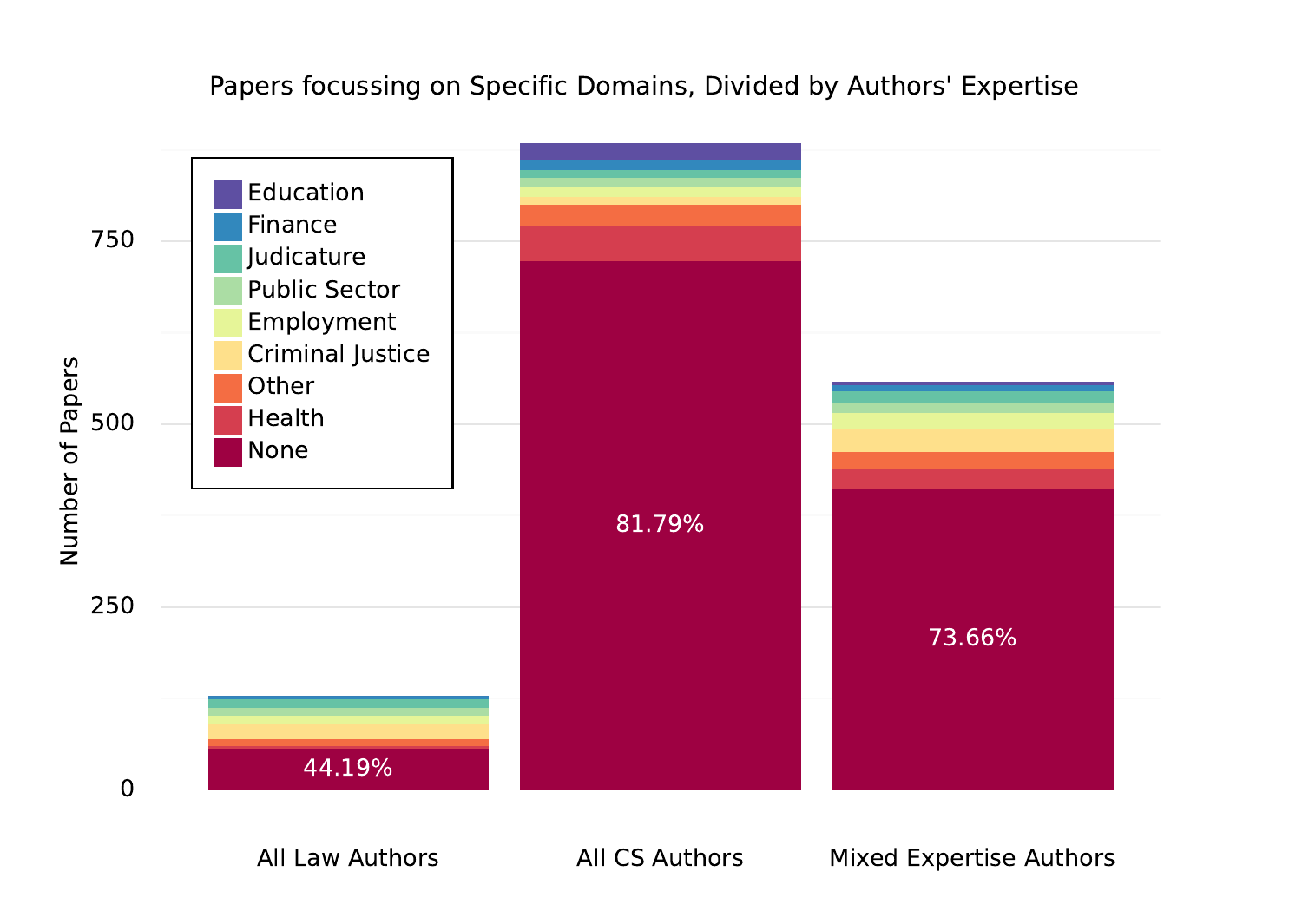}
        \caption{The domain focus of papers split by first author's expertise}
        \label{fig:domain_specific_expertise}
    \end{minipage}
    
    \begin{tikzpicture}[overlay, remember picture]
        \node[inner sep=5pt, draw=black, line width=1pt, text width=0.9\linewidth, align=center, font=\small] at ($(current page.north) + (0,-1.5)$) {%
            \textbf{Common papers in ``Other" category}: Policing (12), Social Media (8), Sharing Economy (7), Advertisement (6), Insurance (3)};
    \end{tikzpicture}
    
    \caption{Over the years papers have become more domain-specific and authors from a legal- background are most likely to write domain-specific papers. The most discussed domains are health, criminal justice and employment}
    \label{fig:approach_specific_info}
\end{figure*}

Looking at Figure \ref{fig:domain_specific_over_years}, we observe a rising trend in the number of domain-specific papers over the years. Whereas in 2016 only \~12\% of papers looked at algorithmic fairness through a domain-specific lens, this has risen to 28\% by 2023. The most prominent domains revolve around health, criminal justice, judicature and employment. Perhaps unsurprisingly, the rising interest in these domains coincides with case studies within those spheres that have gained public attention. For instance, in 2016 ProPublica published their article on the infamous COMPAS case, an algorithmic risk assessment tool that displayed racial biases against African Americans \cite{compas}. In the years following the publication, there is a notable rise in papers addressing fairness in criminal risk assessment. Similarly, we observe an increased interest in fairness in the employment sector after information was released in 2018 about a recruitment tool that Amazon scrapped because of its sexist preference towards male candidates \cite{reuter}. 
As we will argue in the next paragraphs, domain-specific approaches open up doors to not just view algorithmic fairness as a general problem, but approach the topic with awareness of the unique challenges in each domain. This holds both when focussing on algorithmic fairness from a legal and technological point of view. 

When inspecting Figure \ref{fig:domain_specific_expertise} it is striking how legal authors are much more likely to take this approach to algorithmic fairness than Computer Scientists. A common theme that is touched upon by them, is the adequacy of existing laws to address changes brought by the ubiquitous use of AI systems in different domains. 
For example, Hertza examines the regulatory landscape in the United States on credit lending, focusing on the Fair Credit Reporting Act (FCRA) and the Equal Credit Opportunities Act (ECOA) \cite{hertza2018fighting}. He argues that these laws are inadequate in safeguarding the rights of credit consumers in light of the increasing reliance on big data and advanced algorithmic systems for lending decisions. For example, the FCRA gives consumers the right to access their credit report records, consisting of information about their loan history, on which credit decisions were traditionally based. However, given that the FCRA was enacted in the 1970s, it did not account for the type of third-party data that banks increasingly use to make their decisions, such as lenders' social media profiles or web browsing history. 
Lacking the right to access this information and understanding how algorithms utilize it, makes it impossible for consumers to challenge algorithmic decisions and assess their fairness. To make up for these gaps in the legislation, Hertza proposes the adoption of the EU General Data Protection Regulation (GDPR) for reforming consumer credit regulation in the US. Because the GDPR is an industry-agnostic framework, it gives individuals the right to access any personal information being processed about them, not just the information on their credit history, that comes from financial institutions.
Studies like these highlight the advantage of domain-specific approaches when discussing algorithmic fairness from a legal perspective: as many domains come with their own set of laws, only specific contributions can give insights into their adequacy and invite researchers to challenge them. 

Figure \ref{fig:domain_specific_expertise} shows us that Computer Scientists are less likely to take domain-specific approaches. Still, when investigating some of their works, it becomes apparent why specific contibutions are needed to understand the technological challenges within different sectors.
Take for instance Pena et al.'s paper set in the employment domain, exploring the type of algorithms typically used to analyse resumes or other professional profiles (e.g., LinkedIn data) in a hiring context. Different from the typical data in other domains, resumes are usually multimodal, as they consist of structured data (e.g., standardized formats to display a person's educational history), unstructured text (e.g., personal biographies) and even images (e.g. profile pictures). Consequentially, automatized solutions for hiring decisions are also multimodal, meaning that one or more multiple models are built to analyze the different data types and base decisions on them. While biases in text-processing or computer vision models have been studied in isolation, the combination of these models and how this combination contributes to new discriminatory biases is less well studied.  
Another employment-specific study by Rhea et al. even showed how in a hiring setting, simple changes, like whether a resume is processed as raw text or in a PDF file, can change the output of such decision-making systems \cite{rhea2022resume}. We argue that domain-specific approaches are much more likely to reveal problems like these, as they encourage researchers to consider the input data and algorithmic systems that are already in use, rather than making generic assumptions about them.

A final argument for more domain-specific approaches is that they can foster collaborations with interdisciplinary researchers, industry and public institutions, allowing for more in-depth and realistic analyses of unfair practices. Take for instance the study by Elzayhn et al. \cite{elzayn2023measuring}, which is a collaboration between computer scientists, economic- and legal researchers, as well as employees of the US Office of Tax Analysis. They analyse a real-life dataset of taxpayers in the US and - taking domain knowledge about the Tax Payment System into account - analyze if tax audit rates (that partly depend on algorithmic decisions) differ for black and non-black taxpayers. In working with realistic data, the researchers have to deal with challenges often ignored in algorithmic fairness literature, e.g. how to conduct an audit when information about sensitive attributes is not available, but needs to be accurately inferred from the data.
Further, by collaborating with the Tax Analysis Office they identify possibilities in reducing the found racial impact, while accounting for their budget and time constraints.
Again, this paper forms a contrast to more generic work on algorithmic discrimination, where computer scientists often work in isolation of the institutions using algorithmic systems \cite{veale2018fairness}.  In those papers, researchers also commonly use benchmarking datasets for testing their algorithms, which are publicly available datasets, meant to standardize how algorithmic performance is assessed. Though these datasets have their merits, researchers have warned about their quality and the extent to which they can mirror realistic industry use cases \cite{ding2021retiring}.
Additionally, only using benchmark datasets can increase the risk of overgeneralizing results obtained from them. Hence, working with domain-specific data that comes from interdisciplinary/industrial collaborations can provide more realistic views on the suitability of AI technologies aimed at addressing algorithmic biases.

To conclude this section, we believe that generic work has been and can still be useful to lay the foundation for algorithmic fairness, but that having more domain-specific case studies will help in tackling more realistic challenges. We have observed a clear trend towards researchers publishing more domain-specific papers, however, as Figure \ref{fig:domain_specific_over_years} shows, many papers remain generic and others revolve around similar domains like health and criminal justice. This may come at the risk of ignoring the risk of algorithmic discrimination in other domains, such as policing, insurance or sharing economy platforms. Hence, broadening the scope of the research field and keeping up to date with the diverse industries/institutions using algorithmic systems, will be essential to reveal the technological challenges and legislation gaps specific to each domain and create tailor-made solutions for them.

\subsection{RQ2: Making Diverse (Intersectional) Demographic Groups the Focus of the Research} \label{sec_results_demographics}

\begin{figure*}[t]
    \centering
    \begin{minipage}[b]{0.52\textwidth}
        \centering
        \includegraphics[width=\textwidth]{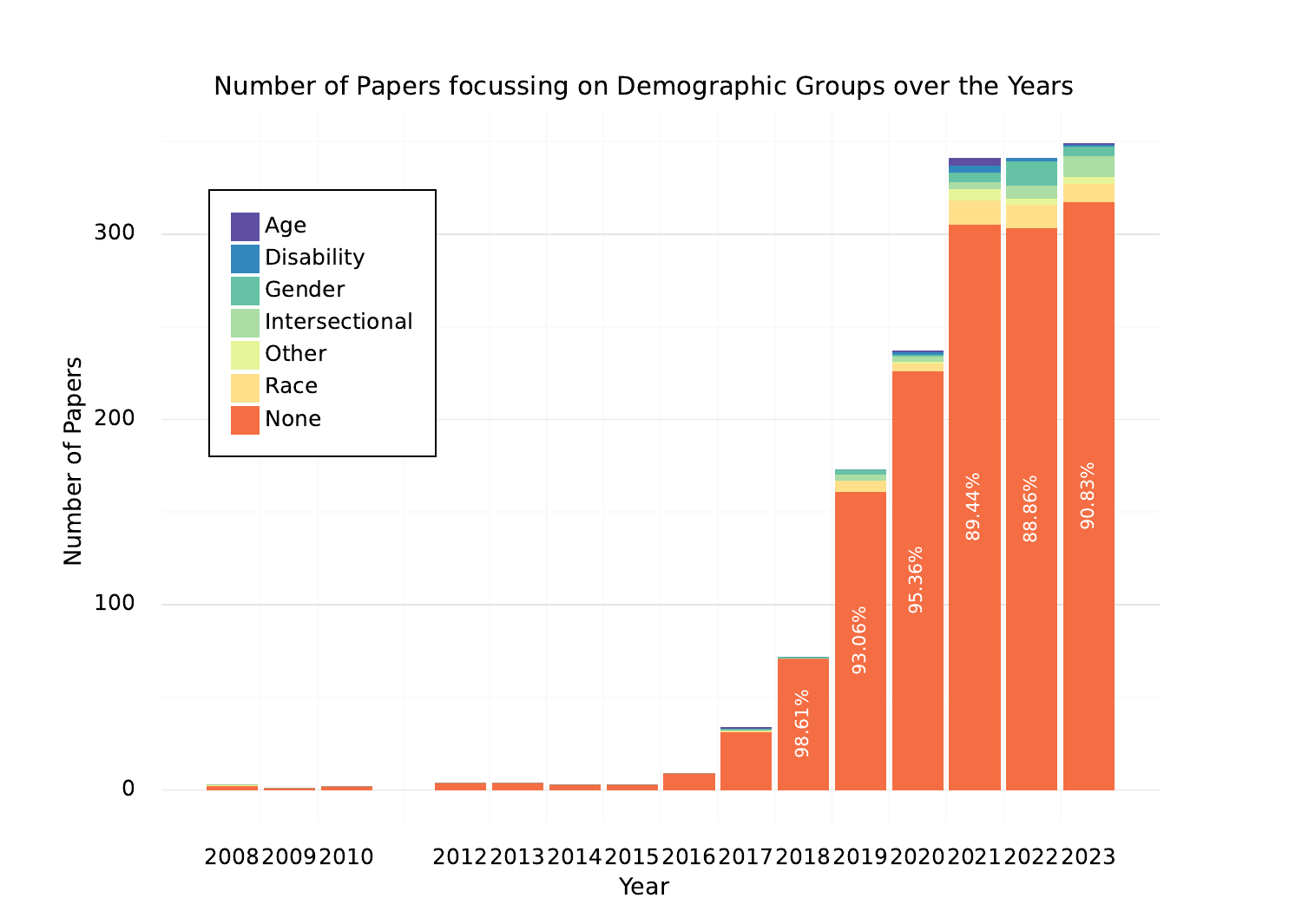}
        \caption{The demographic focus of papers over the years}
        \label{fig:demographic_specific_over_years}
    \end{minipage}
    \hfill
    \begin{minipage}[b]{0.45\textwidth}
        \centering
        \includegraphics[width=\textwidth]{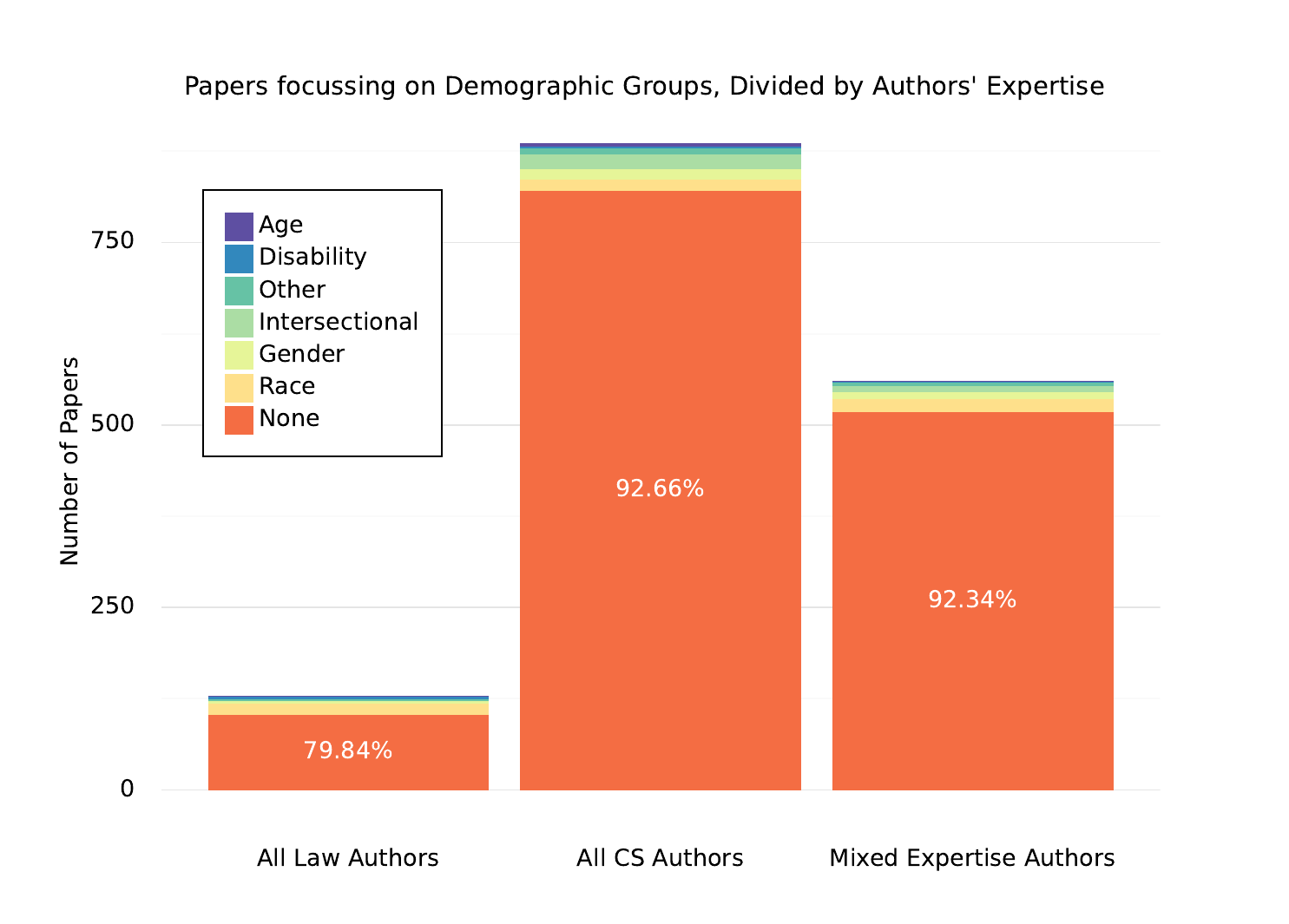}
        \caption{The demographic focus of papers split by first author's expertise}
        \label{fig:demographic_expertise}
    \end{minipage}
    
    \begin{tikzpicture}[overlay, remember picture]
        \node[inner sep=5pt, draw=black, line width=1pt, text width=0.9\linewidth, align=center, font=\small] at ($(current page.north) + (0,-1.5)$) {%
            \textbf{Common papers in ``Other" category}: Economic (6), Nationality (5), Political Opinion (2), Sexuality (2), Religion (1)};
    \end{tikzpicture}
    
    \caption{Most literature tackles algorithmic fairness from a generic perspective, not taking the harms faced by different demographic groups into account.}
    \label{fig:demographic_specific_info}
\end{figure*}

Over the past few years, there has been a slight trend towards publishing more papers that focus on specific demographic groups (e.g., based on race or sex) rather than tackling algorithmic fairness from a generic perspective (see Figure \ref{fig:demographic_specific_over_years}. In 2023, around 10\% of the papers made some demographic group a focus of their work, compared to only 2-6\% in 2017-2019. The most prominent categories that papers focus on are race and gender. Also, noticeably, over the years more papers focused on fairness for intersectional groups. Whereas the first two papers on intersectional discrimination appeared only in 2019, in 2022 and 2023, a combined number of 18 papers have focused on this topic. 

Similarly, when it comes to focusing on domains when studying algorithmic fairness, we believe that focusing on demographic groups comes with the advantage of accounting for the historical context and the social dynamics behind discriminatory practices. This viewpoint echoes the argument presented by Hu \& Kohler-Hausmann in their paper on algorithmic gender discrimination (\citeyear{10.1145/3351095.3375674}). Using the example of a decision-making system for college admissions, they highlight how viewing gender as one physical feature in isolation from other attributes, does not do justice to the broader societal implications that come with it. For instance, in college admissions, there are differences across genders in what college programs they apply for or how people's gender shape their opportunities in life. Hence, any fairness considerations made in such a setting should not just consider questions like ``are admission ratios equally distributed across sexes?", but also consider why women are less likely to apply for science departments or how societal expectations shaped their past educational and extracurricular activities. Only when recognizing what (algorithmic) fairness would mean in the light of these demographic-specific considerations, meaningful steps can be taken to tackle historic inequalities.

While it is encouraging, that some more papers have taken these demographic-specific approaches over the years, the relatively strong focus on gender and race comes with the risk of ignoring other groups that are targets for discrimination.
For instance, over 15 years we only found 8 papers focusing on algorithmic discrimination faced by people with disabilities. While one might argue that general research on algorithmic discrimination is also applicable to ableism, some papers argue why more specialised research is necessary \cite{buyl2022tackling, binns2021could}. 
One argument is that the range of different disabilities is much broader than the range of other sensitive attributes. People with different types of disabilities can be affected by algorithmic systems in vastly different ways. To illustrate, disabilities can range from physical impairments (e.g. being in a wheelchair) to medical conditions (e.g. having cancer), to vision, hearing or cognitive impairments and to psychological conditions (e.g. having depression) \cite{binns2021could}. 
From a technological point of view, Buyl et al. use the example of job recruitment to point out how these distinct categories of disabilities affect algorithms differently (\citeyear{buyl2022tackling}): a person with visual impairments, for instance, may need more time on an automated recruitment test, lowering their chances of making it to the next round of a selection procedure. A person with a history of psychological/medical conditions may not have this problem but may instead have bigger gaps on their CV that may be penalised by an algorithm. Lastly, automated video analysis software, used, e.g. for job interviews, may perform okay on either of both groups but not on people with speech impairments. 
The same paper then discusses the idea of ``reasonable accommodation" as a possible technical solution to address these problems: if algorithmic systems have information on the type of disability of any individual, they can be designed to accommodate each of them. For instance, automated video analysis software could be designed to process sign language to accommodate people with speech impairments. Additionally, algorithms for analysing CVs could be designed to not penalize career gaps if a job applicant has a history of medical conditions.
While these are reasonable adjustments from a technological perspective, contributions from a legal perspective point out how difficult it may be to gather data about peoples' disabilities, as people might prefer not to disclose this information, for fear of it being abused or lack of discretion in handling the data \cite{buyl2022tackling, binns2021could}. A paper by Binns \& Kirkham (\citeyear{binns2021could}), therefore, explores the role of data protection and equality law, in ensuring algorithmic fairness for disabled people, while simultaneously protecting their privacy.
For instance, they highlight how data protection laws (e.g., the GDPR) allow institutions to collect ``special category data" (including information about persons' disability status) if they have an appropriate lawful basis for wanting to process this data. Further, they emphasize how these laws can create a safer and more trustworthy environment around sharing personal data, as they define clear boundaries regarding how the data should be used and with which parties it can be shared. Hence, ensuring strict enforcement of these laws can increase peoples' willingness to share sensitive information and ensure that this information is only used to provide ``reasonable accommodation", as mentioned earlier.

The example papers surrounding algorithmic fairness for disabled people illustrate the importance of delving into specific demographic groups to gain a clearer understanding of how they are affected by algorithmic systems. By outlining both technologically- and legally-driven research papers, we emphasize how expertise from both disciplines is needed to find realistic solutions for the addressed challenges. Inspecting Figure 6, this is especially a call for Computer Science researchers to adopt such demographic-specific approaches, as they are less likely to do so than their legal counterparts. Specifically, only around 7\% of Computer Scientists make specific demographic groups the main focus of their research, while this ratio is 14\% higher for Law experts.
Further, we emphasize again, how the research on algorithmic fairness needs to broaden its scope and include various demographic groups that go beyond just the race and gender of people. As Figure \ref{fig:demographic_specific_info} points out, there are still many demographic features that are barely considered in current research efforts, posing the risk of overlooking the harms faced by diverse and intersectional communities.

\subsection{RQ3: Moving beyond Classification} \label{sec_results_approach}
\begin{figure*}[t]
    \centering
    \begin{minipage}[b]{0.52\textwidth}
        \centering
        \includegraphics[width=\textwidth]{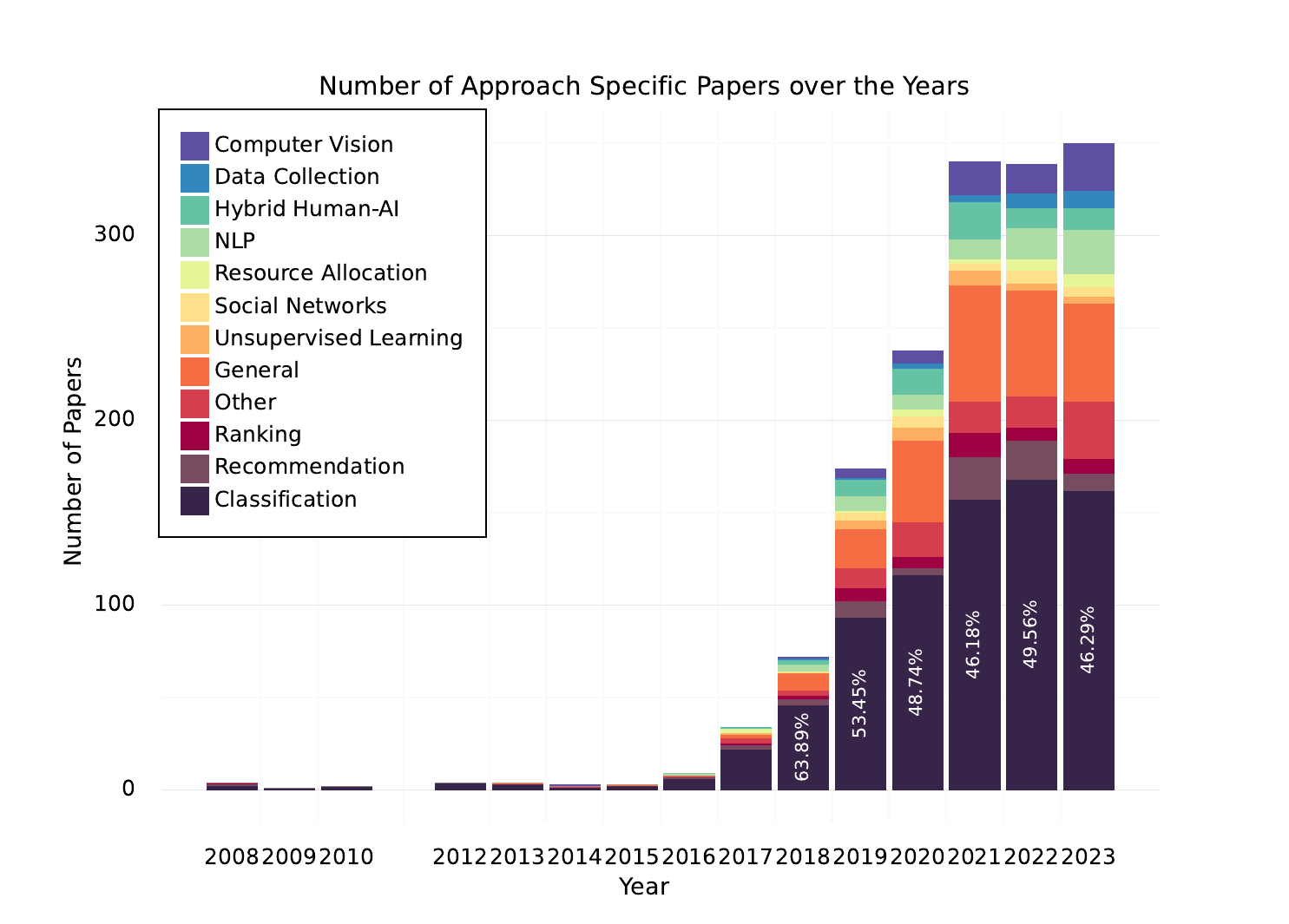}
        \caption{The technological focus of papers over the years}
        \label{fig:technology_specific_over_years}
    \end{minipage}
    \hfill
    \begin{minipage}[b]{0.45\textwidth}
        \centering
        \includegraphics[width=\textwidth]{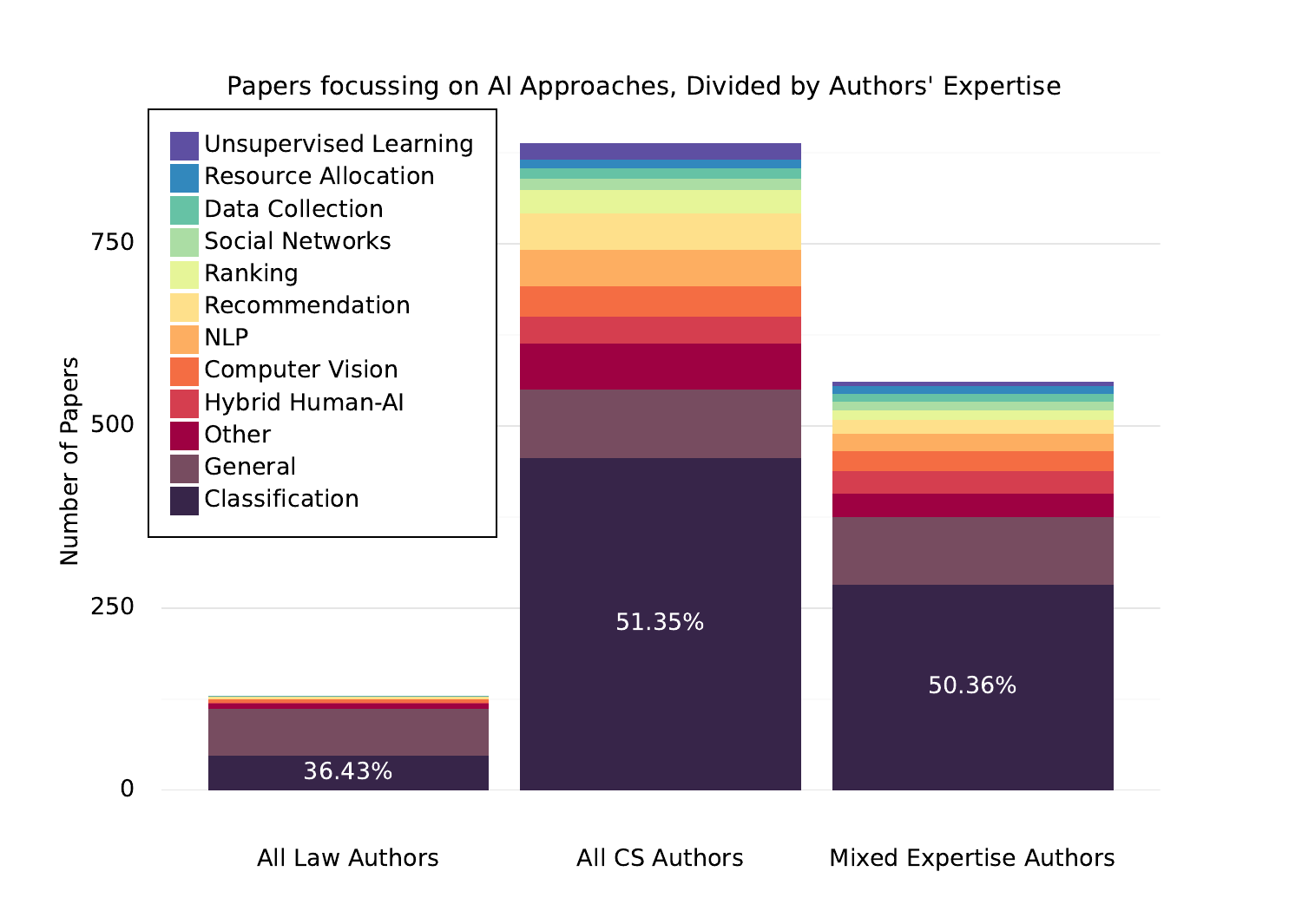}
        \caption{The technological focus of papers split by first author's expertise}
        \label{fig:technology_expertise}
    \end{minipage}
    
    \begin{tikzpicture}[overlay, remember picture]
        \node[inner sep=5pt, draw=black, line width=1pt, text width=0.9\linewidth, align=center, font=\small] at ($(current page.north) + (0,-1.5)$) {%
            \textbf{Common papers in ``Other" category}: Generative AI (13), Federated Learning (11), Speech Recognition (10), 
            Price Discrimination (8), Representation Learning (8), Regression (7), 
            Internet of Things (6)};
    \end{tikzpicture}
    
    \caption{Classification remains the prime technological focus of studies on algorithmic fairness. Computer Science research is slowly getting more diverse in their addressed technologies.}
    \label{fig:technology_specific_info}
\end{figure*}

In Figure \ref{fig:technology_specific_over_years}, we display how focuses on different approaches have developed over the years. What is striking, yet unsurprising, is that over the years ``Classification" remains the most discussed technology regarding algorithmic fairness. While on the one hand, this may be a result of our generic search query, which did not include specific terms like ``Clustering” or ``Speech Recognition”, it undoubtedly reflects an already known concern, that researchers often do not look beyond fairness in classification tasks based on tabular data \cite{holstein2019improving, richardson2021towards, chouldechova2020snapshot}. Still, we observe a trend that the range of discussed technologies gets wider and more diverse over the years, as in 2018 still about 64\% of all contributions focussed on classification while by 2023 this has gone down to 46\%. The most heavenly discussed technologies besides classification are computer vision, NLP and recommendation systems.

When inspecting where more diverse discussions on AI technologies come from in Figure \ref{fig:technology_expertise}, we immediately see that authors from a pure Law background are the least diverse, with nearly all their contributions discussing classification tasks or ``AI" as a general phenomenon. Partly, this may result from the generic ways in which AI legislation is phrased. Since technology develops so rapidly and unpredictably, it is impossible to account for all its potential forms. Hence generic guidelines around its usage allow policy writers to encompass more use cases, likely reflected in the scientific literature about these guidelines. 
Still, neglecting the precise shapes that algorithms can take can lead to an incomplete understanding of their usage and substantial gaps in the laws regulating them. This is exemplified in one of the legal contributions from Keunen (\citeyear{keunen2023big}). She investigates data collection practices around tax audits and the extent to which they can be considered ``fishing expeditions". Specifically, she examines their privacy-intruding nature, wherein an excessive amount of taxpayer information is collected and analyzed in the pursuit of detecting fraud, before having sufficient justification for why these taxpayers are targeted as potentially fraudulent and why extra data needs to be collected for them.
In her work, Keunen alludes to various technologies used to collect this data, namely automated web scraping and web crawling algorithms. While she primarily raises privacy concerns related to these practices, it is clear how also from an algorithmic fairness standpoint these techniques can be highly problematic. For instance, another work by Jo \& Gebru, explains how the availability and nature of data that can be crawled from online spaces is influenced by demographic factors (\citeyear{jo2020lessons}), with e.g. younger generations being more represented on the internet than older ones. Consequently, fraud-detection algorithms relying on web-crawled data may disproportionately impact younger groups, as more potentially incriminating data is available about them.
Despite the clear fairness and privacy concerns around web crawling, Keunen points out that their regulation and the extent to which they can be considered ``fishing expeditions" is still unclear: explicit legislation is not available and so far only case law serves as an indication for which data collection practices are prohibited (\citeyear{keunen2023big}). Hence, Keunen's work showcases, how for identifying other gaps in the legislation related to privacy and algorithmic fairness, more legal experts need to dive into specific technologies, rather than primarily focusing on AI as a general problem. Collaborating with Computer Science experts in doing so, will be important to stay on top of the fast-paced development of technology. 

To highlight some of the complex fairness considerations, that Computer Scientists currently make about other non-classification technologies consider the work of Jalal et al. (\citeyear{jalal2021fairness}), who explore image-reconstruction algorithms. These algorithms take low-resolution images as input and try reconstructing them into higher resolutions. In doing so, they are known to be biased. For instance, when low-resolution images of a black person are given as input they are likely to reconstruct it into the image of a white person. The work addresses the intricacies of even defining ``fairness" in such a setting. Unlike classification tasks, where a classifier's decision should be independent of sensitive attributes (e.g., employment decisions should not be influenced by race), fair image reconstruction algorithms must produce outputs that align with the sensitive characteristics in the input. This introduces the challenge of estimating race and other sensitive attributes from images, a task complicated by their non-discrete and highly ambiguous nature. 
Another example of non-trivial fairness issues concerns the use of Generative AI systems. While our scoping review found only 13 papers related to this technology, it seems reasonable to assume that this number will rise, given the popularity of ChatGPT, DallE, and other generative systems. Venkit et al. (\citeyear{narayanan2023unmasking}) are some of the few authors exemplifying the fairness issues arising through these systems, examining how text generation models exhibit different sentiments and toxicity levels depending on the nationalities they are prompted to write about. For example, when prompted to write about Irish people, human annotators perceived the articles to be mostly benign and generic, while texts about Tunisian people were rated to be much less positive and more focused on negative events in the country. How such texts can perpetuate harmful stereotypes and how to restrict these models are still largely unexplored questions. The topic is complicated considerably by the seemingly infinite topics these systems can be prompted to write about and all the ways a chatbot-human interaction could unfold. Hence, for just defining what it means for such a huge system to be fair, more technological and legal research is necessary.

While it lies outside the scope of this paper to discuss the fairness concerns arising in all other kinds of algorithmic systems, it should be clear that they can go far beyond the matters addressed in the typical classification setting. As technology advances rapidly and various algorithmic systems become more prevalent among the public, it is clear that researchers should make an effort to keep up with this development and extend their focus beyond the conventional realms.\looseness=-1

\subsection{RQ4: Considering Global Perspectives}
To analyse the geographical context in which algorithmic fairness was discussed, we both examined the authors' affiliation countries as well as the geographic focus in papers' content. For the former analysis, we considered the first authors as the primary contributors.

\subsubsection{Authors' Affiliation}
\begin{figure*}[t]
    \centering
    \includegraphics[width=0.65\textwidth]{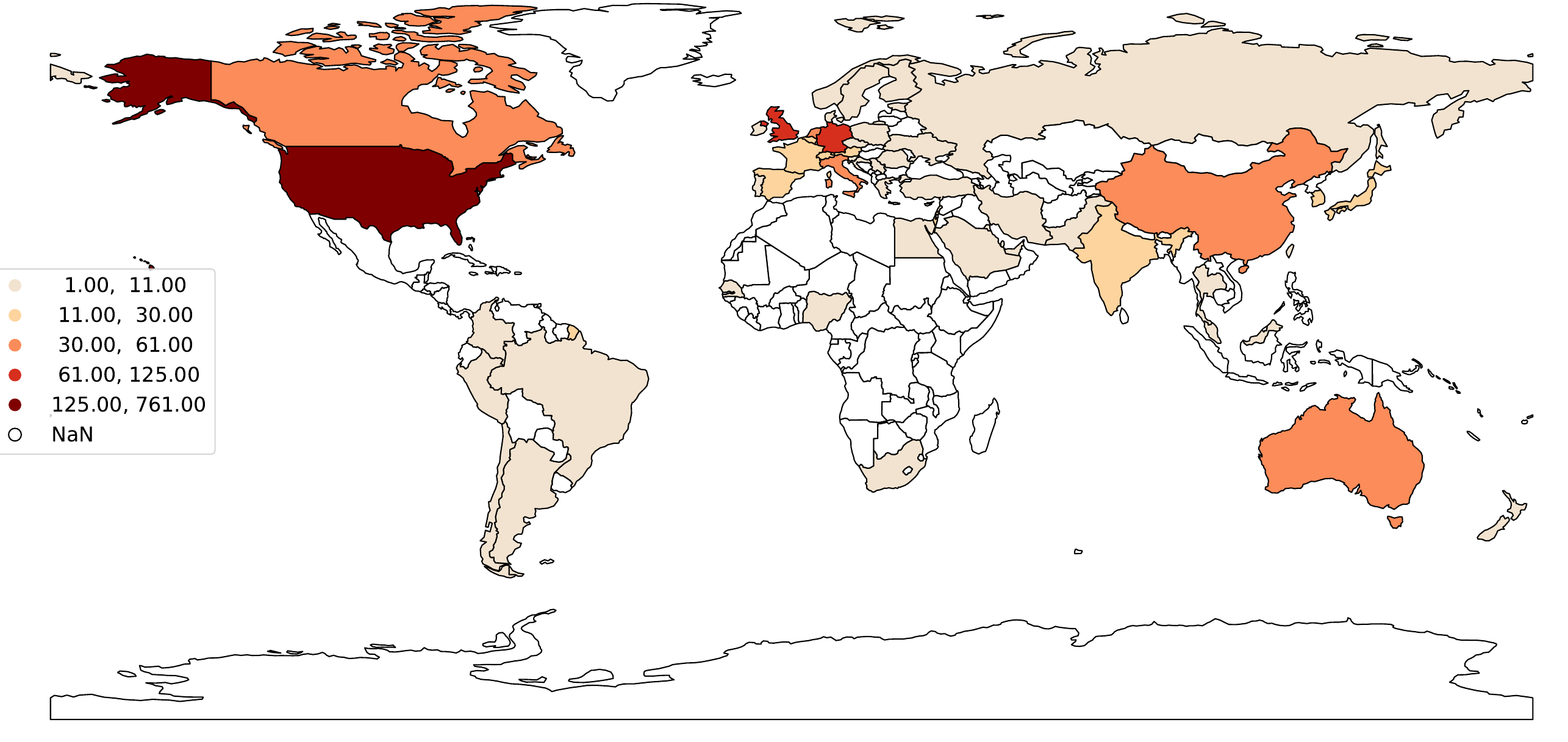}
    \caption{A world map displaying the affiliation countries of each paper's first author }
    \label{fig:first_author_map}
\end{figure*}

In Figure \ref{fig:first_author_map}, we display a geographical heatmap, displaying the number of papers divided by each paper's first author's affiliation country. At first sight, it is evident that most contributions come from authors affiliated with institutes in North America and Europe, and papers from authors affiliated with other countries are quite sparse.  To further investigate this trend, we classified each first author's affiliation country according to whether it belongs to a Global North or Global South country\footnote{We used UNCTAD’s classification in which Global North is understood as countries in Europe and Nothern America; and including Israel, Japan, Australia and New Zealand. The Global South consists of countries in Africa, Asia, South America and the Caribbean. \url{https://unctadstat.unctad.org/EN/Classifications.html}}. In Figure \ref{fig:gn_gs_over_years}, we display how this geographical context of the first authors has developed over the years.
From this Figure, we see that consistently most contributions come from authors hailing from the Global North.
While this predominant presence persists, a noteworthy shift is discernible over the years, with an almost 20\% uptick in contributions from the Global South institutions in 2023, signalling a gradual re-balancing compared to earlier years where only about 5\% of contributions come from the Global South.
As we will see in the following section, this shift is crucial for challenging the Northern-centric perspective within AI research.

\begin{table*}[!htbp]
    \begin{subfigure}[t]{0.58\textwidth}
        \centering
        \includegraphics[width=\textwidth]{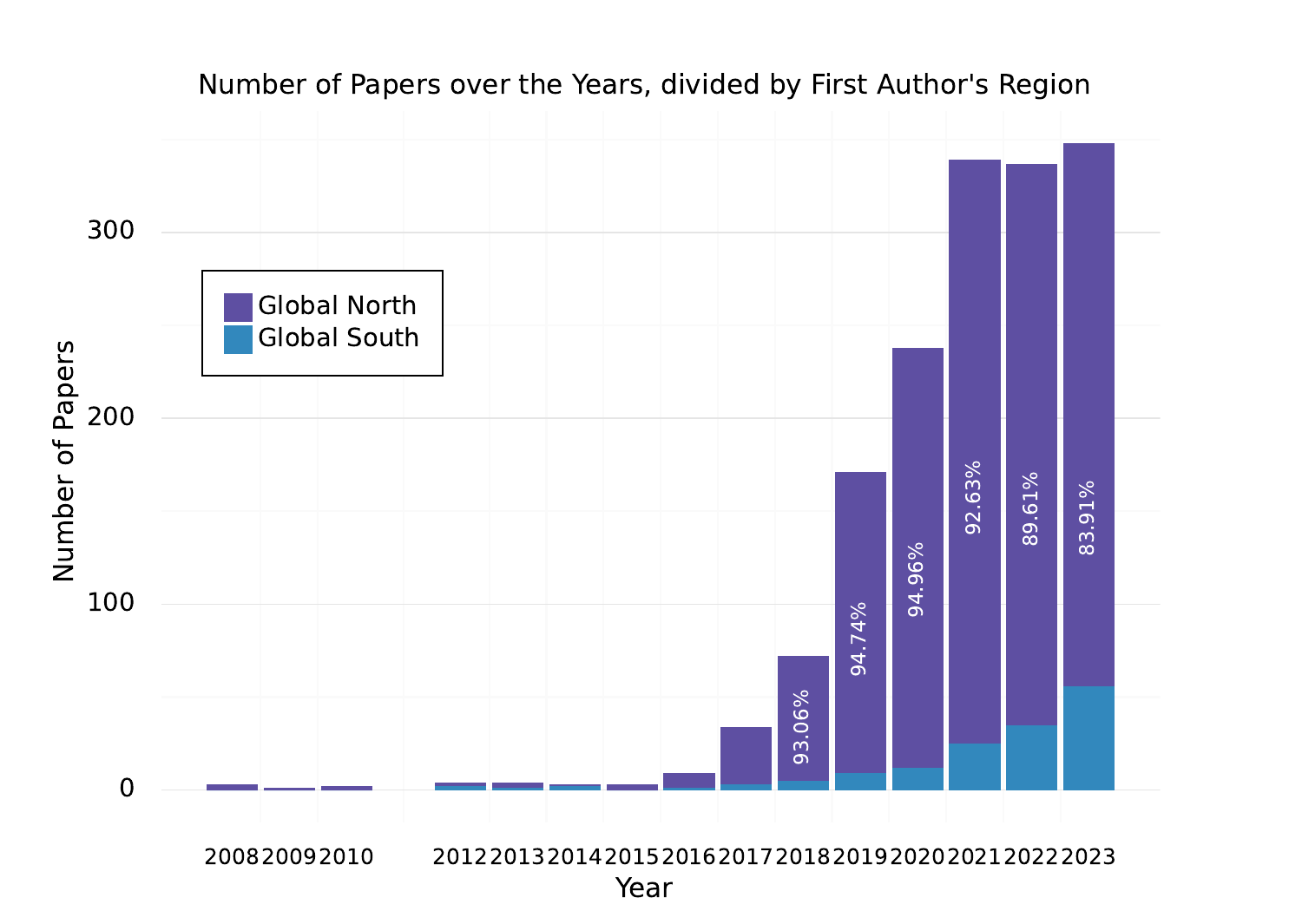}
        \caption{Though the majority of papers come from authors affiliated in the global north, contributions from papers from global southern affiliations are rising}
        \label{fig:gn_gs_over_years}
    \end{subfigure}
    \hfill
    \centering
    \begin{subtable}[t]{0.4\textwidth}
        \centering
        \raisebox{1.2\height}{
        \scalebox{0.8}{\begin{tabular}{l|l}
        \textbf{Country/Regional Focus} & \textbf{\# Papers} \\ \hline
        United States & 82 \\ \hline
        Europe & 60 \\ \hline
        United Kingdom & 12 \\ \hline
        China & 9 \\ \hline
        India & 7 \\ \hline
        Australia & 6 \\ \hline
        Canada & 4 \\ \hline
        Brazil, Netherlands, Germany, Italy, Africa & 3 \\ \hline
        \begin{tabular}[c]{@{}l@{}}Spain, Singapore, France, Austria, Russia, \\ Chile, Global South\end{tabular} & 2 \\ \hline
        \begin{tabular}[c]{@{}l@{}}New Zealand, Switzerland, Uruguay, Mexico,\\ South America, Bangladesh, South Korea, \\ Maldives, Vietnam, Philippines, Japan, Asia, \\ Israel, United Arab Emirates, Nigeria\end{tabular} & 1
        \end{tabular}}}
        \caption{Number of times different countries/geographical regions were made the focus of a papers' content}        \label{tab:geo_focus}
    \end{subtable}
    
\end{table*}

\subsubsection{Geographical Focus in Papers' Content}
Next to the first authors’ affiliation country, we analysed the papers' geographic focus, as estimated by them mentioning any specific country/region name in their title or their abstract. In Table \ref{tab:geo_focus} we display the results of this analysis. 

Intriguingly, the results unveil a similar representation gap, as we have found upon examining the authors' affiliation, with most papers concentrating on countries in the Global North.
There could be several reasons for the under-representation of work from/about the Global South, especially those from Africa:
\begin{itemize}
    \item Databases may not systematically include publications from the Global South, hinting at access challenges or a predilection for regional databases.
    \item The search query methodology, requiring specification of the Global South country in the title, may inadvertently limit the breadth of results.
\end{itemize}

These, among other reasons, such as the limited resources in research institutions in the Global South, language barriers, and lack of engagement with literature from the South, contribute to the underrepresentation of certain geographical regions\cite{nakamura2023three}.

Nevertheless, as already mentioned, the past years have observed a rise in both papers coming from non-northern institutions, as well as a rise of papers concentrating on algorithmic fairness in global southern countries. Notable examples include studies on Predictive Policing in New Delhi \cite{newdelhipredictivepolicing}, Early Prediction of At-Risk Students in Uruguay \cite{uruguayriskstudents}, and discussions on Algorithmic Fairness in China and Brazil \cite{wang2020black, discriminationbrazil}. These papers delve into the intricacies of AI applications within diverse cultural, economic, and legal contexts, emphasising the need for nuanced considerations in algorithmic development. 
The paper “The Algorithmic Imprint ” by Eshan et al. (\citeyear{ehsan2022algorithmic}) provides an especially clear example of why these kinds of considerations are necessary, and why it is important to have more diverse and inclusive voices in the narrative about algorithmic fairness.
The paper discusses the grading algorithm developed to predict students’ A Level results when the exams could not be administered due to the Covid19 pandemic. Though there were many unfairness complaints about the algorithm’s predictions (and they were ultimately discarded) they turned out to be especially unfair towards students from commonwealth schools outside of the UK, in which A-Levels are also the primary form of examination. By focussing on the specific case of Bangladeshi schools, the authors find how UK-based assumptions throughout the algorithm’s development, can explain the disparity in predictions. One example is, that the grade predictions were based on performance in mock exams, assuming that good performance on a mock exam is predictive of a good performance on the real one. While intuitively this might make sense, this assumption neglects the learning culture in Bangladesh where much more emphasis is put on final examination and mock exams are not a common part of the curriculum. To have some data to work with, Bangladeshi students were forced to take some hurriedly set up tests, which they were not used to and had little time to prepare for. Needless to say, grade predictions based on this type of data, did not reflect students' real capabilities.
Another flaw in the design of the algorithm was the decision to base grade predictions on the historical performance of the student's school. If such data were not available, international averages were used instead. This proved especially problematic for Bangladeshi schools, which were less likely to possess (digital) records of historical performance. Consequently, predictions frequently leaned on the international averages, even though these were lower than the (unrecorded) actual historical performances.
These are only a few of many examples, of how a lack of cultural considerations led to an algorithm, that was ultimately more unfair to some geographic groups than others. 

Having additional papers adopting a cultural- and geographic-specific approach can contribute to a more diverse and comprehensive understanding of algorithmic fairness, shedding light on various perspectives and mitigating unfairness across different regions. 
In addition to specific case studies, we also found several papers contributing to a more global discourse on algorithmic discrimination. For instance, Amugongo et al. (\citeyear{sambala2020ubuntu}) examine fairness from a philosophical standpoint, exploring how African-based ``Ubuntu ethics" can enrich discussions about the essence of fairness. Another example is Nwafor's paper (\citeyear{nwafor2021ai}), which delves into the policies and laws from global southern countries concerning AI systems. Studies like these will be essential to make sure that legislation outside Europe and the US is ready for upcoming technological developments.
In her paper Nwafor also advocates for a diverse representation in AI's design, development, deployment, and governance. Neglecting to engage marginalised communities in AI's development, leads to technological innovation being based on a a narrow slice of the world, lacking a comprehensive analysis of diverse global groups. Integrating more diverse perspectives not only enhances our understanding of algorithmic fairness but also emphasizes the importance of cross-cultural learning to create more inclusive and equitable AI systems.\looseness=-1

\section{Discussion \& Conclusion}
In this paper, we have presented a scoping review of the current literature on algorithmic fairness. We selected and annotated 1570 papers to examine the evolution of the field in terms of their domain-, demographic-, and technological focus and their interdisciplinary nature, while also inspecting the geographical context in which the research takes place.

We acknowledge two major limitations in our analysis. First, we used a basic search query to collect papers for our dataset by using general terms such as "machine learning" and "AI". However, this approach did not account for more specialised terms regarding papers technological approach, their domain or demographic focus, which may have caused us to miss valuable contributions within those areas.\looseness=-1

The second limitation concerns our manual annotation process, which we based on the papers' titles and abstracts, rather than their full text. While both should give a good reflection on the paper's main topic, some important nuances might have been missed.

Despite these limitations, our results still provide a valuable overview of the current research landscape surrounding algorithmic fairness, in particular the trending topics and the gaps within the field. 
Our analysis shows that over the years, research has started focusing on more specific and a wider variety of topics in terms of the addressed technologies, domains and demographic groups. This stands in contrast to early work in the field, which discussed algorithmic fairness concerns solely in classification tasks, without questioning domain-specific challenges or the harms different demographic groups might face. Through highlighting some papers, we have made a case for why more specialised research is necessary, both from a legal and technological point of view, as non-specific approaches come at the risk of ignoring the algorithmic systems that are actually used by companies and the unique considerations that go into tackling their discriminatory behaviour.  

Finally, we examined the geographical context of ongoing research, by analysing authors' affiliations and the papers' geographical focus. While the trend is slowly changing, most papers come from global north countries and focus on the algorithmic development and regulations there. Through some case studies, we have emphasized how a lack of diverse cultural considerations in developing algorithms, can lead to severe discriminatory results depending on where they are applied. Therefore, an inclusive approach is necessary to comprehend the broader implications of algorithmic fairness in distinct contexts and how to address these.

\bibliography{aaai24}

\end{document}